\documentclass{webofc}
\usepackage[varg]{txfonts}   
\begin{document}
\title{Pierre Auger Observatory and Super Heavy Dark Matter}

\author{
\firstname{Roberto} \lastname{Aloisio}\inst{1,2}\fnsep\thanks{roberto.aloisio@gssi.it}
\lastname{for the Pierre Auger Collaboration}\inst{3}\fnsep\thanks{spokespersons@auger.org. \\ 
\indent  A complete list of authors can be found at  \url{https://www.auger.org/archive/authors_2022_09.html}
}}

\institute{
	Gran Sasso Science Institute, viale F. Crispi 7, 67100 L'Aquila, Italy 
\and
	INFN - Laboratori Nazionali Gran Sasso, via G. Acitelli 22, 67100 Assergi (AQ), Italy
\and
	Observatorio Pierre Auger, Av. San Mart\'in Norte 304, 5613 Malarg\"ue, Argentina
         }

\abstract{
We briefly discuss the connection of the Pierre Auger Observatory data with a large class of dark matter models based on the early universe generation of super heavy particles, their role in the solution of the dark matter problem, highlighting the remarkable constraining capabilities of the Auger observations.  
}
\maketitle
\section{Introduction}
\label{intro}

Ultra-high energy cosmic rays (UHECR) are the most energetic particles ever observed with energies larger than $10^{20}$ eV. These extreme energies, as high as $10^{11}$ GeV, eleven orders of magnitude above the proton mass and "only" eight below the Planck mass, are a unique workbench to probe new ideas, models and theories beyond the Standard Model (SM) of particle physics, which show their effects at energies much larger than those ever obtained, or obtainable in the future, in accelerator experiments. This is the case of theories with Lorentz invariance violations \cite{PierreAuger:2021tog,Addazi:2021xuf,Aloisio:2000cm,Aloisio:2002ed} or models of Dark Matter (DM) with super heavy particles \cite{PierreAuger:2022ibr,PierreAuger:2022wzk,Aloisio:2006yi,Guepin:2021ljb,Aloisio:2015lva,Aloisio:2007bh}, that connect UHECR observations with the Dark Sector (DS) and the cosmological evolution of the early universe.  

The leading paradigm to explain DM observations is based on the Weakly Interactive Massive Particle (WIMP) hypotheses, a stable DM particle with mass in the range of $0.1\div 10$ TeV \cite{Bergstrom:2000pn}. Searches for WIMPs provided no evidence of a clear candidate and alternative solutions to the DM problem should be considered. An alternative to WIMP models  is represented by the scenarios based on long lived super-heavy relics, that can be produced by several mechanisms taking place during the inflationary phase or just after, in the re-heating phase \cite{PierreAuger:2022ibr,PierreAuger:2022wzk,Aloisio:2006yi,Aloisio:2015lva,Aloisio:2007bh}. Once created in the early universe, the abundance of the long-lived super-heavy particles can evolve to match the DM density observed today, the so-called Super Heavy Dark Matter (SHDM). This conclusion can be drawn under three general hypotheses: (i) SHDM in the early universe never reaches local thermal equilibrium; (ii) SHDM particles have mass $M_X$ of the order of the inflaton mass or higher; and (iii) SHDM particles are long-living particles with a lifetime exceeding the age of the universe, $\tau_X\gg t_0$.

By far the largest experiment devoted to the observation of UHECR is the Pierre Auger Observatory (Auger) in Argentina. UHECR can be observed only indirectly through the detection of their interaction products with the Earth's atmosphere. An UHE particle interacting with a nucleus of the atmosphere produces both hadronic and electromagnetic cascades of particles, collectively called Extensive Air Shower (EAS). In Auger the EAS observation is performed by detecting the EAS particles that reach the ground and by the (coincident) detection of the fluorescence emission produced in the atmosphere, the so-called hybrid events.

As always in the case of cosmic rays, the information we can gather experimentally concerns: energy spectrum, mass composition and arrival direction anisotropy. The Auger observations of UHECR clarified several important facts \cite{Coleman:2022abf,AlvesBatista:2019tlv,Aloisio:2017ooo,Aloisio:2017qoo,PierreAuger:2022uwd}: (i) UHECR are charged particles, with severe upper limits on UHE photon and neutrino fluxes (see later); (ii) the spectrum observed at the Earth shows a slight flattening at energies around $5\times 10^{18}$~eV (called the ankle) with (iii) an instep at $10^{19}$~eV and (iv) a steep suppression at the highest energies; (v) mass composition is dominated by light particles (proton and helium) at energies around $10^{18}$~eV becoming progressively heavier starting from energies around $3\times 10^{18}$~eV, with lacking of light particles toward the highest energies; (vi) at energies larger than $8\times 10^{18}$~eV UHECR show a strong signal of dipole anisotropy in the arrival directions, with $6\%$ amplitude and the phase pointing toward the direction of the galactic anti-center.

In section \ref{models}, we will briefly introduce SHDM models discussing their connection to the Auger observations and the related constraining capabilities. We conclude in section \ref{conclusions}.

\section{Super-heavy dark matter models and constrains}
\label{models}

Recently, using the LHC measurements of the masses of the Higgs boson and Top quark, it has been pointed out that the Higgs potential is stable until very high energy scales $\Lambda_I\simeq 10^{10} \div 10^{12}$ GeV and it might be possible to extrapolate the SM to even higher energies up to the Planck mass $M_P=10^{19}$ GeV \cite{Buttazzo:2013uya}. This fact implies that the mass spectrum of the DS is restricted to ultra-high energies and SHDM could play a role. In the following, we will consider SHDM with masses in the range between $10^{8}$ GeV up to the Planck mass. 

Apart from the gravitational interaction, SHDM can be coupled to ordinary matter also through some super-weak coupling, driven by a high energy scale $\Lambda>\Lambda_I$ as the Grand Unification Scale ($\Lambda_{GUT} \simeq 10^{16}$ GeV). 
\begin{figure}[!h]
\centering
\includegraphics[scale=.228]{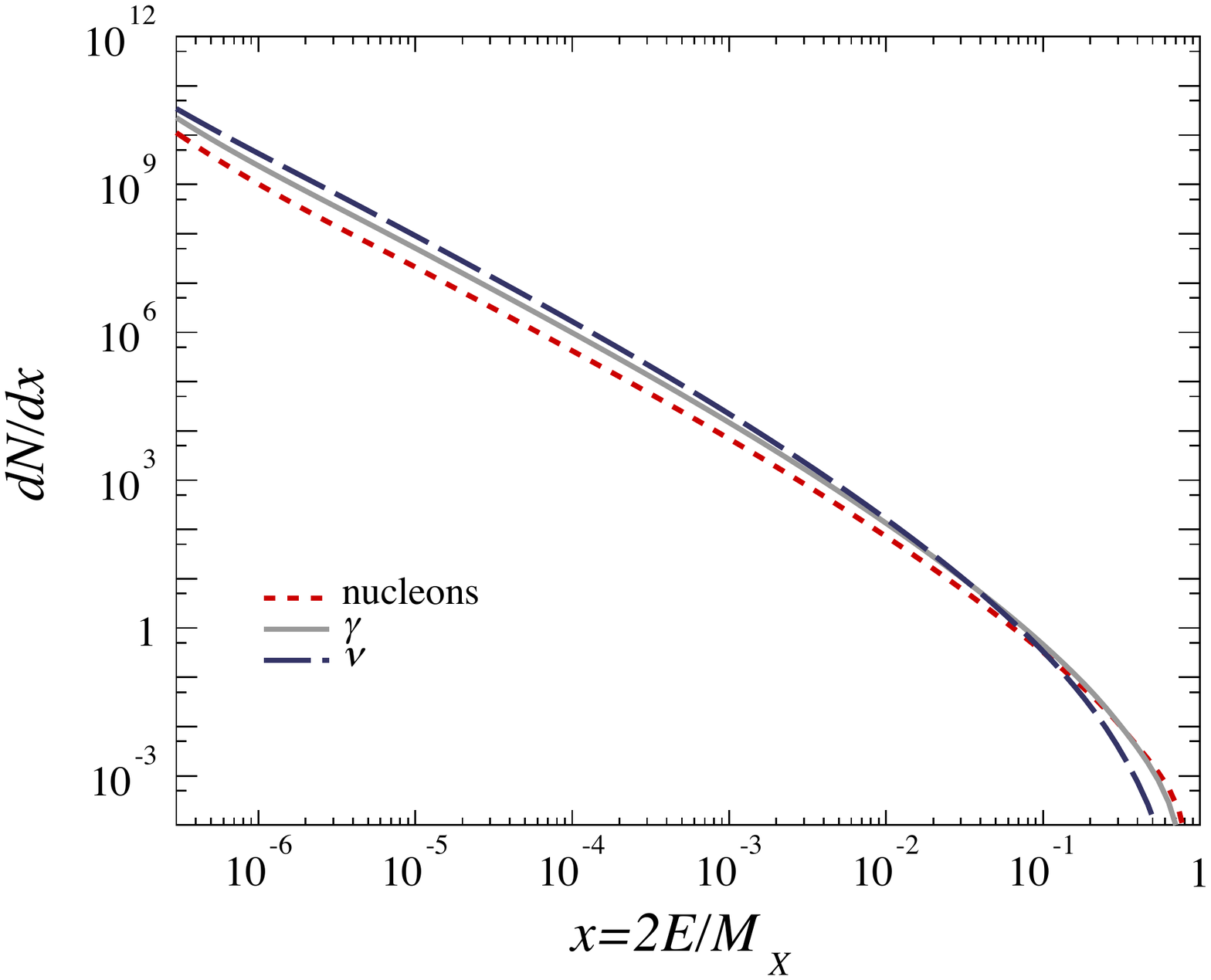}
\includegraphics[scale=.20]{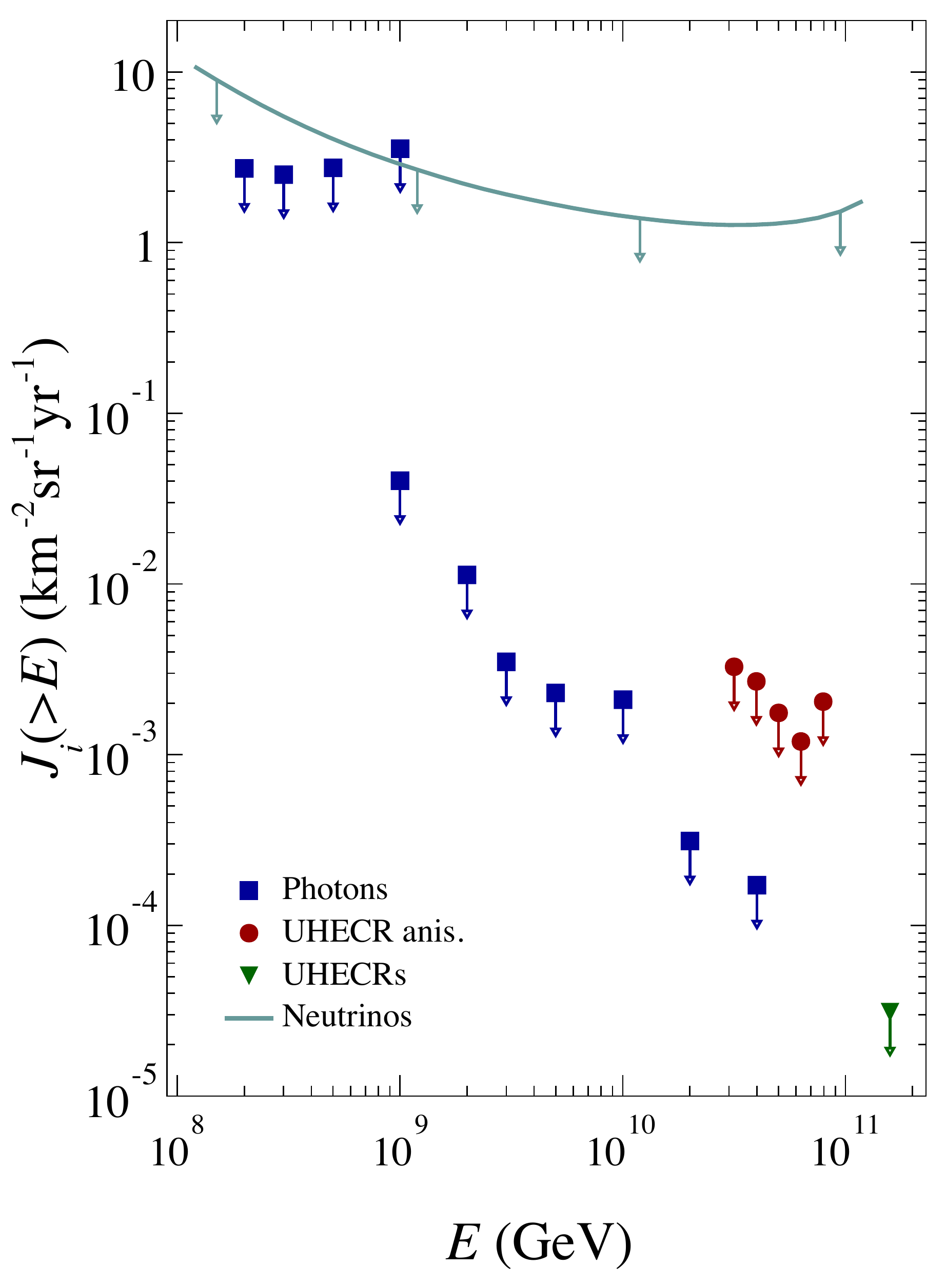}
\caption{ [Left Panel] Distribution function of the SM products in the SHDM decay. [Right Panel] Auger experimental limits on the fluxes of gamma rays and neutrinos. Figures taken from \cite{PierreAuger:2022ibr}.}
\label{fig1}  
\end{figure}
To assure long-lived particles, the interaction SHDM-SM should be suppressed by some power $n$ of the high energy scale $\Lambda$, with a decay SHMD $\rightarrow$ SM characterised by a lifetime $\tau_X\simeq (M_X \alpha_{X\Theta})^{-1} (\Lambda/M_X)^{2n-8}$, being $\alpha_{X\Theta}$ the reduced coupling constant between SHDM and SM particles \cite{PierreAuger:2022ibr}. 
In the case in which SHDM interacts with SM particles only through the gravitational interaction\footnote{Note that, in this case, a SHDM particle can be also called Planckian-Interacting Massive Particle (PIDM).}, assuming the very general case of a DS characterised by its own non-abelian gauge symmetry, SHDM can decay only through non-perturbative effects as the instanton-induced decay \cite{PierreAuger:2022ibr}. In this case, the lifetime of SHDM follows from the corresponding instanton transition amplitude that, being exponentially suppressed, provides long-living particles. Considering the zeroth order contribution, the instanton-induced lifetime can be written as $\tau_X\simeq M_X^{-1} \exp{(4\pi/\alpha_X)}$, being $\alpha_X$ the reduced coupling constant of the hidden gauge interaction in the DS.

Under very general assumptions \cite{PierreAuger:2022ibr,PierreAuger:2022wzk,Aloisio:2006yi,Guepin:2021ljb,Aloisio:2015lva,Aloisio:2007bh}, we can determine the composition and spectra of the standard model particles produced by the SHDM decay. Typical decay products are couples of quark and anti-quark\footnote{For a discussion of alternative (leptonic) decay patterns see \cite{Guepin:2021ljb}.} that, through a cascading process (jets), give rise to neutrinos, gamma rays and nucleons. As follows using the DGLAP equations \cite{Aloisio:2006yi}, these particles exhibit a flat spectrum, that, as shown in the left panel of figure \ref{fig1}, at the relevant energies, can be approximated as $dN/dE \propto E^{-1.9}$, independently of the particle type, with a photon/nucleon ratio of about $\gamma/N\simeq 2\div 3$ and a neutrino nucleon ratio $\nu/N\simeq 3\div 4$, quite independent of the energy range \cite{Aloisio:2006yi}. Therefore, the most constraining limits on the SHDM mass and lifetime are those coming from the (non) observation of UHE photons and, even to a lesser extent, neutrinos. 

Photons with energies larger than $10^{8}$ GeV have an absorption length roughly of the size of our galaxy, increasing up to the scale of the local group ($1\div 10$ Mpc) at energies larger than $10^{10}$ GeV. Moreover, as follows from numerical simulations of structure formation, DM is clustered in galactic halos with an over-density at the level of $10^{5}$ \cite{Aloisio:2006yi}, it implies that only UHE photons from the decay of SHDM in our our own galaxy play a role. The observation of photons with energy larger than $10^{8}$ GeV is affected by the background of cosmogenic photons, i.e. UHE diffuse photons produced by the propagation of UHECR. This background depends on the mass composition of UHECR and on the details of their astrophysical sources, it can be estimated to range (at most) around $10^{-2}$ km$^{-2}$ sr$^{-1}$ yr$^{-1}$ above $10^{8}$ GeV and around $10^{-3.5}$ km$^{-2}$ sr$^{-1}$ yr$^{-1}$ above $10^{9}$ GeV \cite{PierreAuger:2022uwd}. These cosmogenic diffuse fluxes are two orders of magnitude below the sensitivity of Auger and can be safely neglected. The search for UHE photon events in the Auger dataset exploits the well known differences in the EAS initiated by photons or hadrons: EAS generated by photons develop deeper in the atmosphere and are characterised by a reduced number of muons, with a substantial detectable difference in the lateral distribution of EAS particles \cite{PierreAuger:2022uwd}. In the right panel of figure \ref{fig1} we plot the very stringent limits on the photon and neutrino fluxes as extracted from the Auger data (2022 update) in the energy range $10^{8}\div 10^{11}$ GeV \cite{PierreAuger:2022ibr,PierreAuger:2022wzk,PierreAuger:2022uwd}.

\begin{figure}[!h]
\centering
\includegraphics[scale=.22]{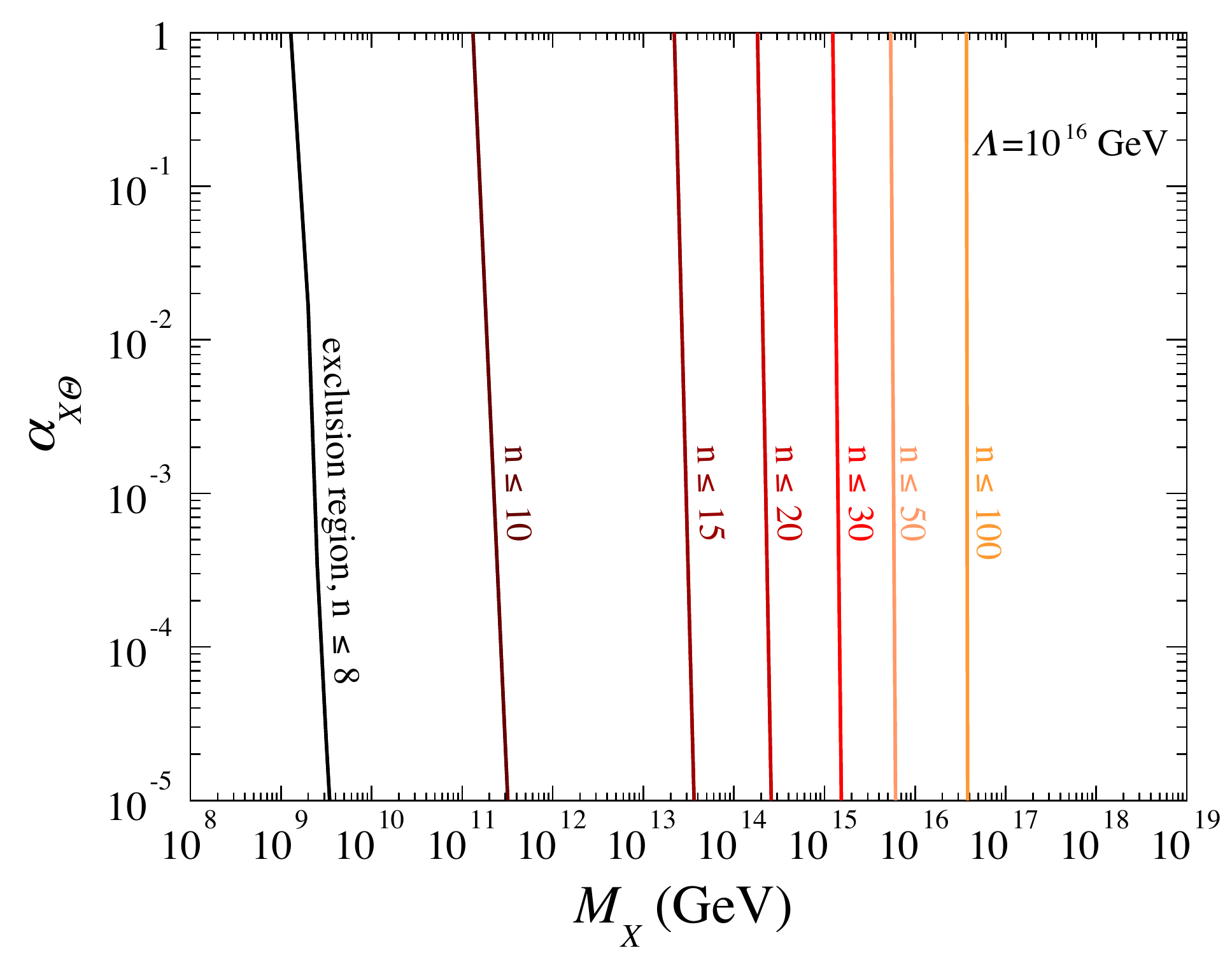}
\includegraphics[scale=.21]{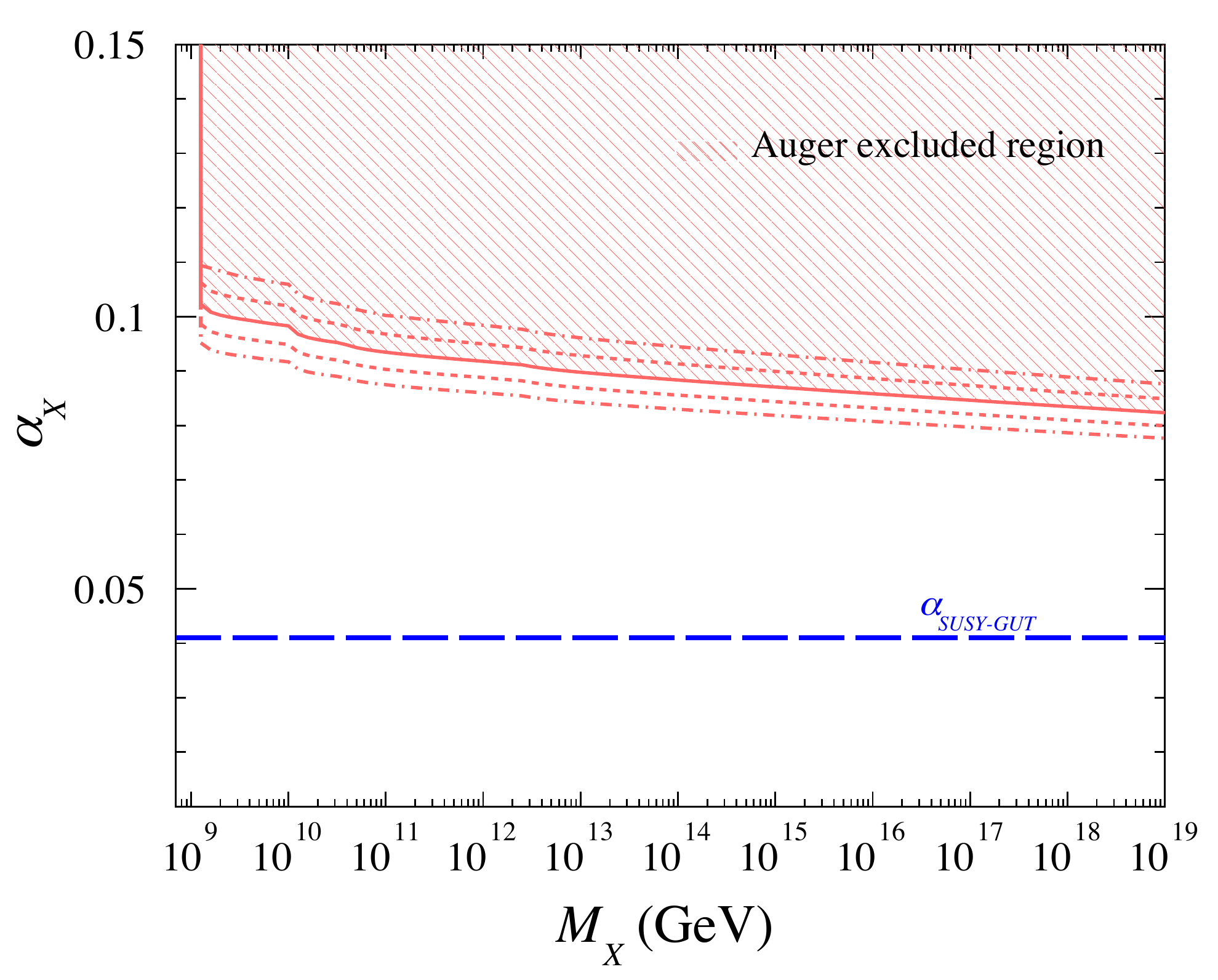}
\includegraphics[scale=.21]{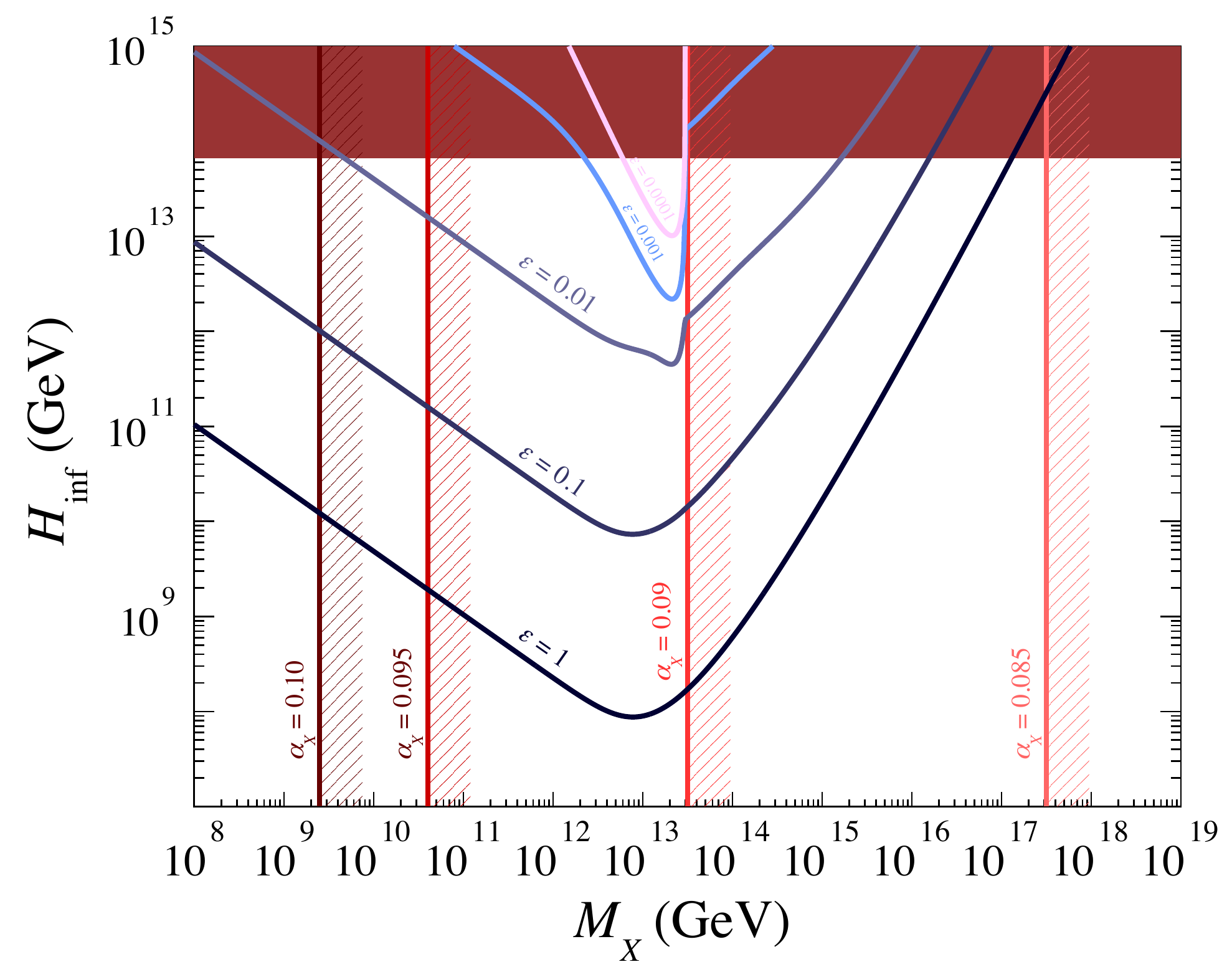}
\caption{Limits fixed by the Auger observations on the SHDM mass $M_X$ and: (left panel) the coupling SHDM-SM $\alpha_{X\Theta}$ for different values of the suppression scale order $n$, (central panel) the coupling of the hidden DS gauge interaction $\alpha_{X}$ in the case of the instanton-induced decay, (right panel) the universe expansion rate at the end of inflation $H_{inf}$. In the right panel, we plot the limits for different values of the reheating efficiency $\epsilon$ (see text). Figures taken from \cite{PierreAuger:2022ibr}.}
\label{fig2}
\end{figure}

Assuming a DM density in our galaxy as in the Navarro-Frenk-White profile \cite{Navarro:1995iw}, we can determine the expected flux of UHE photons produced by the decay of SHDM. Requiring that this flux does not exceed the Auger experimental limits, it is possible to constrain the mass and lifetime of SHDM. In the left and central panels of figure \ref{fig2}, we plot the limits on $M_X$ and the coupling SHDM-SM $\alpha_{X\Theta}$, in the case of direct coupling (left panel), and the coupling of the hidden DS gauge interaction $\alpha_{X}$, in the case of the instanton-induced decay (central panel). Assuming the class of models based on SHDM creation during the reheating phase, the limits on $M_X$ and $\tau_X$ can be rewritten in terms of the cosmological parameters characterising the generation mechanism of SHDM. In the right panel of figure \ref{fig2}, we plot the limits coming from Auger data on the age of the universe at the end of inflation $H_{inf}^{-1}$ for different values of the reheating efficiency $\epsilon=(\Gamma_\phi/H_{inf})^{1/2}$, being $\Gamma_\phi=g_\phi^2 M_\phi/8\pi$ the inflaton decaying amplitude into SHDM ($M_\phi$ inflaton mass; $g_\phi$ inflaton coupling with SHDM) \cite{PierreAuger:2022ibr,PierreAuger:2022wzk}.

\section{Conclusions}
\label{conclusions} 

We conclude by stating the importance of Auger observations on UHE photons and neutrinos, as they enable to constrain models of new physics connected to the grand unification scale and the reheating phase of the universe. This is the case of SHDM that, as discussed above, being a viable alternative to the WIMP paradigm for DM, can be tested only through UHECR observations, eventually connected with cosmological observations.

\end{document}